\documentclass[twocolumn,showpacs,preprintnumbers,amsmath,amssymb,prl,superscriptaddress]{revtex4}
\usepackage[]{graphicx}% Include figure files
\usepackage{dcolumn}% Align table columns on decimal point
\usepackage{bm}% bold math
\usepackage{color}

\newcommand{\NA}{N_{\rm A}}
\newcommand{\eSi}{{^{28}\rm Si}}
\newcommand{\dd}{{d_{220}}}

\begin{document}

\preprint{IAC/NAh 28Si d220 1.0}

\title{Measurement of the \{220\} lattice-plane spacing of a $^{28}$Si crystal}% Force line breaks with \\

\author{E.\ Massa}
\author{G.\ Mana}
\affiliation{INRIM -- Istituto Nazionale di Ricerca Metrologica, strada delle cacce 91, 10135 Torino Italy}%
\author{U.\ Kuetgens}
\affiliation{PTB -- Physikalisch-Technische Bundesanstalt, Bundesallee 100, 38116 Braunschweig Germany}%
\author{L.\ Ferroglio}
\affiliation{INRIM -- Istituto Nazionale di Ricerca Metrologica, strada delle cacce 91, 10135 Torino Italy}%
\date{\today}

\begin{abstract}
The spacing of the \{220\} lattice planes of a $^{28}$Si crystal was measured by combined x-ray and optical interferometry to a $3.5\times 10^{-9}$ relative accuracy. The result is $d_{220}=(192014712.67 \pm 0.67)$ am, at 20.0 $^\circ$C and 0 Pa. This value is greater by $(1.9464 \pm 0.0067)\times 10^{-9} \dd$ than the spacing in natural Si, a difference which confirms quantum mechanics calculations. Subsequently, this crystal has been used to determine the Avogadro constant by counting the Si atoms, a key step towards a realization of the mass unit based on a conventional value of the Planck or the Avogadro constants.
\end{abstract}

% WASO04 impurity free, d220 = 192014340.00(1.10) am @ 20.0 °C and 0 Pa

\pacs{06.20.-f, 06.20.Jr, 61.05.cp, 07.60.Ly}%
% 06.20.-f Metrology
% 06.20.Jr Determination of fundamental constants
% 06.30.Bp Spatial dimensions
% 61.05.cp X-ray diffraction
% 07.60.Ly Interferometers
\keywords{metrology, fundamental constants, x-ray interferometry}

\maketitle

We participated in a project ended with an accurate measurement of the Avogadro constant, $\NA$, by counting the atoms in a silicon crystal highly enriched with $\eSi$ \cite{Avogadro,NA:PRL}. This could make it possible to realize the mass unit on the base of the mass of an atom \cite{kg-1}. Since the molar Planck constant is known via the measurement of the Rydberg constant, this measurement will also lead to a value of the Planck constant, $h$, and, in turn, to a kilogram realization based on a conventional value of $h$.

The measurement equation is $N_{\rm A} = nM/(\rho a_0^3)$ \cite{Deslattes}, where $M$ is the molar mass, $\rho$ the density, $a_0^3$ the cubic unit cell volume, $a_0$ the lattice parameter, and $n$ the number of atoms per unit cell. The mass of the Pt-Ir kilogram-prototype may have changed by about 50 $\mu$g since its manufacturing, in 1889 \cite{kg-2}. Owing to the smallness of this drift, the relative uncertainty of the $\NA$ determination must not exceed $2\times 10^{-8}$.

The crystal production started in 2004, with the isotope enrichment of the SiF$_4$ gas by the Central Design Bureau of Machine Building  in St.\ Petersburg; subsequently, after conversion of the enriched gas into SiH$_4$, a polycrystal was grown by chemical vapor deposition at the Institute of Chemistry of High-Purity Substances of the Russian Academy of Sciences in Nizhny-Novgorod. Eventually, a 5 kg $\eSi$ boule was grown and purified by application of the float-zone technique by the Leibniz-Institut f\"ur Kristallz\"uchtung (Berlin) in 2007 \cite{28Si}. The pulling speed was so chosen as to reduce the self-interstitial concentration; no doping by nitrogen was applied. The growth axis was [100].

\begin{figure}\centering
\includegraphics[width=8.5cm]{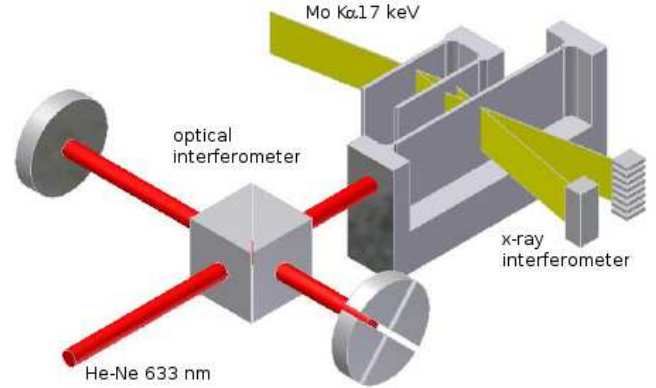}
\caption{The combined X-ray and optical interferometer.} \label{XINT}
\end{figure}

The lattice parameter $a_0$ is one of the quantities required to determine $\NA$; the most accurate way to measure it is by means of combined x-ray and optical interferometry \cite{Bonse,Deslattes:d220,d220:PTB,d220:IMGC}. Owing to the many contributions to the $\NA$ uncertainty, the uncertainty of this measurement should be reduced to $3\times 10^{-9}a_0$. Therefore, we extended the measurement capabilities of an x-ray interferometer to many centimeters and manufactured an interferometer with unusually long analyzer. The measured value was also connected to the results of measurements of the lattice parameter of the natural Si reference-crystals of the PTB and the INRIM \cite{mo*4,waso42a,waso04}.

\paragraph{Experimental apparatus.} The combined x-ray and optical interferometer is shown in Fig.\ \ref{XINT}. It consists of three crystal slabs, 1.20 mm thick, so cut that the \{220\} planes are orthogonal to the crystal surfaces. X-rays from a Mo K$_\alpha$ source are split by the first crystal and recombined, via two transmission crystals, by the third, called the analyzer. When the analyzer is moved along a direction orthogonal to the \{220\} planes, a periodic variation in the transmitted and diffracted x-ray intensities is observed, the period being the diffracting-plane spacing. The analyzer movement is an extremely difficult task; it requires nanoradian attitude-control and picometer vibration and position controls.

The analyzer embeds front and rear mirrors, so that its displacement and rotation can be measured by optical interferometry; the necessary picometer and nanoradian resolutions are achieved by polarization encoding and phase modulation. The symmetric analyzer shape allows it to be reversed; measurements can be carried out with the x rays crossing the crystal in both directions so that surface effects, as well as others systematic effects, can be investigated.

\begin{figure}\centering
\includegraphics[width=6.3cm]{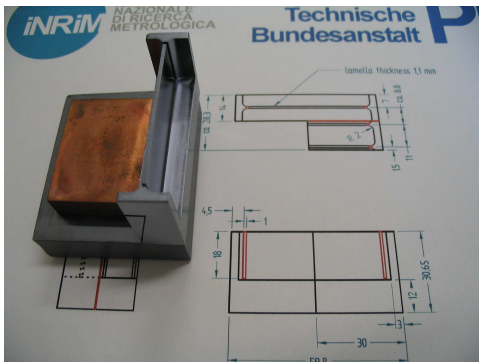}
\caption{Photograph of the $\eSi$ analyzer on its silicon support. The crystal temperature is measured inside the copper block.} \label{photo}
\end{figure}

The measurement equation is $a_0 = \sqrt{8}\dd = \sqrt{8}m \lambda/(2n)$, where $\dd$ is the spacing of the analyzer diffracting-planes and $n$ is the number of x-ray fringes observed in a displacement of $m$ optical fringes having period $\lambda/2$. The laser source operates in single-mode and its frequency is stabilized against that of a transition of the $^{127}$I$_2$ molecule. This ensures the calibration of the optical interferometer with a negligible uncertainty. To eliminate the adverse influence of the refractive index of air and to ensure millikelvin temperature uniformity and stability, the experiment is carried out in a thermo-vacuum chamber.

In  practice, $\dd$ is determined by comparing the unknown period of the x-ray fringes against the known period of the optical ones. This is done by measuring the x-ray fringe phase at the start and end of a sequence of increasing displacements $m\lambda/2$, where $m=$ 1, 10, 100, 1000, 3000, and 30000. To determine the fringe phase, the least-squares method is applied; the input data are about 300 samples of six fringes, with a 100 ms integration time and a sample duration of 30 s. Each $\dd$ measurement is the average of about 9 values collected in measurement cycles during which the analyzer is repeatedly moved back and forth along the selected displacement. The visibility of the x-ray fringes approached 50\% with a mean brilliance of 500 counts/s/mm$^2$.

To cope with the highly demanding request of accuracy, the INRIM extended the crystal-displacement capabilities of its combined x-ray and optical interferometer to 5 cm. This magnification made more numerous effects visible and reproducible. In addition, it allowed wider crystal parts to be surveyed, thus increasing confidence in the crystal perfection and in the mean $\dd$ value.

\begin{figure}\centering
\includegraphics[width=8.6cm]{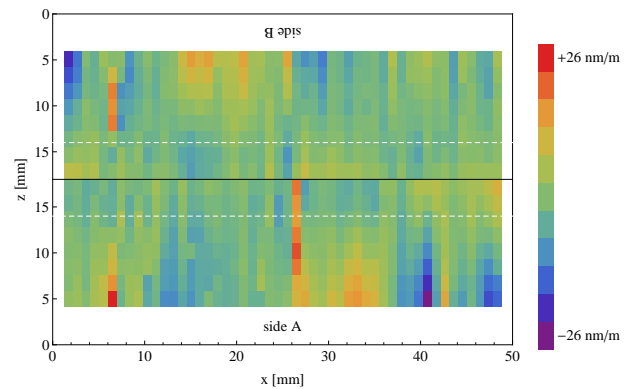}
\caption{Survey of the diffracting-plane strain; obverse (side A) and reverse (side B) analyzer-configurations. The lattice parameter has been measured along the dashed line, which is the virtual extension of the laser beam.}\label{map}
\end{figure}

\paragraph{X-ray interferometer.} In order to exploit the large displacement capability, the PTB manufactured an interferometer with an unusually long analyzer, which is shown in Fig.\ \ref{photo}. To nullify the error due to the different ways the displacement is measured by the x-ray and optical interferometers, by projecting it on the normals to the mirror and diffracting planes, the front and rear mirrors were accurately polished parallel to the \{220\} planes, to within 10 $\mu$rad in the worst case. The analyzer movement was also sensed and controlled on-line to bisect the angle between these normals. Eventually, the measured value was corrected for the remaining imperfections.

\begin{figure}[b]\centering
\includegraphics[width=8.5cm]{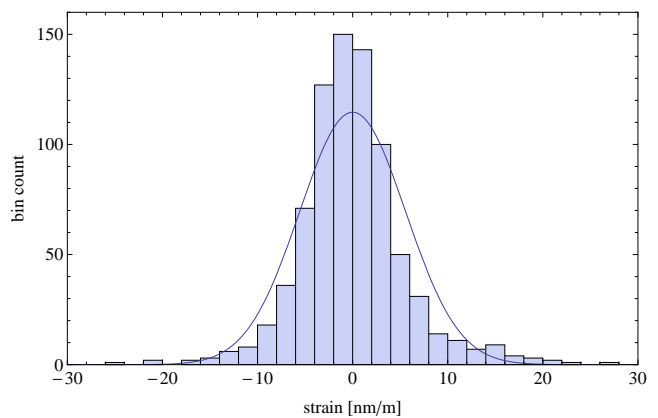}
\caption{Histogram of the observed strains; the standard deviation is 5.6 nm/m. The solid line is the best fit normal distribution.}\label{histo}
\end{figure}

The analyzer has been tested for the lattice perfection. To this end, $\dd$ measurements were made over 50 subsequent crystal slabs, 1 mm wide; x-ray intensities were recorded by means of a multianode photomultiplier having a vertical pile of eight NaI(Tl) scintillator crystals. Hence, the data were processed to obtain a $\dd$ value in each of the $50 \times 8$ pixels, $(1\times 1.8)$ mm$^2$ each. The relative $\dd$ variations are shown in Fig.\ \ref{map}; Fig.\ \ref{histo} shows the relevant histogram. The obverse and reverse configurations correspond to the interferometer crystals mounted as they were in the boule or in a reversed arrangment.

\begin{figure}\centering
\includegraphics[width=8.5cm]{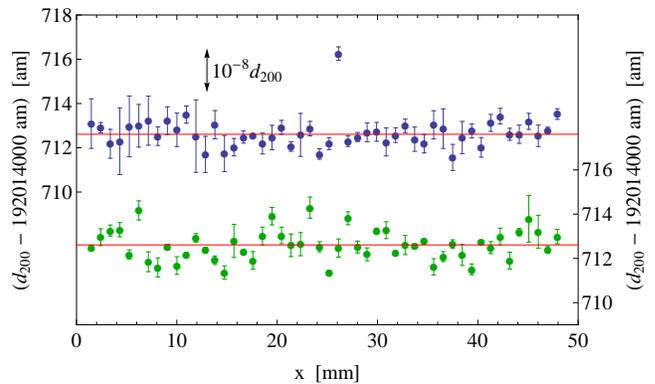}
\caption{Diffracting-plane spacing values measured along the lines indicated in Fig.\ \ref{map}; obverse (top) and reverse (bottom) analyzer-configurations.}\label{d220}
\end{figure}

\paragraph{Results.} Figure \ref{d220} shows the diffracting-plane spacing values measured along the lines displayed in Fig.\ \ref{map}, which indicate the laser beam propagation. These values are insensitive to the Abbe error. This error is due to parasitic rotations combined with an offset between the trajectories of the points sensed by the x rays (indicated by the line in Fig.\ \ref{map}) and the laser beam (indicated by the impinging point on the front mirror). In addition to interpolating the measurement results to obtain the $\dd$ values in the points having a zero Abbe offset, parasitic rotations were sensed and nullified on-line. The values in Figs.\ \ref{map} and \ref{d220} are the averages of different surveys carried out between November 2009 and July 2010; the error bars indicate the standard deviations of the data.

The top panel of Fig.\ \ref{d220} shows an outlier, which survived to the averaging; it is also visible in the corresponding panel of Fig.\ \ref{map}. As shown in the bottom panel, it disappears when the analyzer is reversed. Therefore, we inferred that it does not indicate an actual $\dd$ variation, but an apparent one. An explanation is a surface effect; the phase difference between the interfering x rays, which is the basic measured quantity, records the crystal surface, as well as any extraneous material on it. Though this phase-contrast image is weaker by orders of magnitude than the lattice image, at the sensitivity level we are operating it could affect the measurement result. Because of the the cupric-ion etching used to remove the grinding damage, the analyzer surfaces, though flat on the average, are quite rough; they display a texture with 100 $\mu$m periodicity and 10 $\mu$m peak-to-valley amplitude \cite{waso04}. Since we surmized that the observed outlier originates from this texture or from a residual contaminations, we removed this datum from any subsequent analysis.

Apart from this outlier, none of the measured $\dd$ values was exactly re-observed in different surveys. Additionally, no overlapping is possible between the $\dd$ values obtained with the observe and reverse mounting of the analyzer. Among the many crystals we examined \cite{mo*4,waso42a,waso04}, this is the first one displaying a $\dd$ profile having flatness and smoothness only limited by the residual scattering of the data. Since this scattering is larger than expected from the noises of the x-ray and optical interferometers, investigations are under way to squeeze further the measurement resolution.

\begin{figure}\centering
\includegraphics[width=8.5cm]{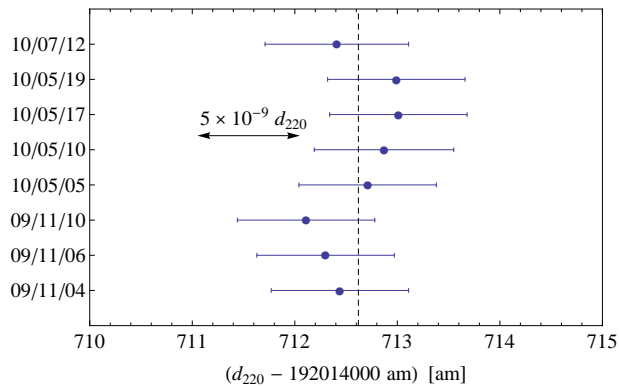}
\caption{Mean of the diffracting-plane spacing values measured along the lines indicated in Fig.\ \ref{map}; the dashed line is the average. Measurements from 09/11/04 to 10/05/05 were made with the obverse analyzer, the following ones with the reverse analyzer. On 10/07/12, the divergence of the laser beam was changed. Up to 10/05/05 the measured values were corrected for the platform tilt; afterwards, it was nullified by a counter-rotation.}\label{comparison}
\end{figure}

To acquire the mean lattice parameter of the analyzer, the $\dd$ values measured along the lines displayed in Fig.\ \ref{map} were averaged and the corrections listed in the Table \ref{table:1} were taken into account. Measurements started in November 2009 and were repeated in May and July 2010; the results are compared in Fig.\ \ref{comparison}. An exemplar error budget is given in Table \ref{table:1}. A detailed analysis of corrections and of error contributions can be found in \cite{mo*4,waso42a,waso04}, which give also the results of test measurements aimed at establishing a firm link with the $\dd$ values of natural Si crystals used as input data for the calculation of a self-consisted set of values of physical constants \cite{CODATA}. Here we draw attention to a couple of points which deserve particular attention.

In the first place, it must be noted that the largest correction is due to the diffraction of the laser beam. At this level of accuracy, the relation $\lambda=c/\nu$ (the symbols having the usual meanings) is not valid; energy disperses outside the region in which it would be expected to remain, wavefronts bend, and their spacing varies from one point to an other and it is different from the wavelength of a plane wave. Fortunately, the relevant correction depends only on beam divergence; not on specific characteristics of the beam, such as the intensity profile \cite{fourier}. In July 2010, a preliminary checking of this result has been made by replacing the fiber collimator with a new one. The beam divergence is the width of its angular power-spectrum and was measured with the aid of a converging lens. It changed from 0.170(8) mrad -- the value relevant to all the previous measurements -- to 0.189(9) mrad; consequently, the correction changed from the $7.26(65) \times 10^{-9} \dd$ value given in Table \ref{table:1} to $8.92(72) \times 10^{-9} \dd$. More important, a beam-profile change was detected, but, as the Fig.\ \ref{comparison} shows, the result of the July 7-th measurement is consistent with the previous values.

In the second place, a new systematic effect came into evidence. To avoid power dissipation inside the vacuum chamber and temperature gradients, the pointing system of the laser beam, including the electro-optical crystal for phase modulation, are outside the chamber, stiffly clamped to it. However, to cut off the vibrations of the vacuum chamber, the experiment platform inside the chamber rests on three O-rings. Owing to their limited stiffness, the hysteresis, and to the mass of the analyzer carriage (about 2 kg), the analyzer displacement causes random misalignment and a systematic tilt of about 200 nrad/mm between the laser beam and the optical interferometer. In this way, in addition to increasing noise and scattering of the measured values, the variation of the lengths optical paths of the laser beam through the interferometer optics causes a systematic error. A separate experiment quantified it in $(1.37\pm 1.20) \times 10^{-9}\dd$. At present, we either corrected the measured values (from 09/11/04 to 10/05/05) or applied a feedforward compensation of the tilt by counter-rotating the platform (from 10/05/10 onwards). Future activity is aimed at measuring and controlling on-line the platform tilt, as well as at reducing the sensitivity of the optical interferometer to jitter and tilt of the laser beam.

\paragraph{Conclusions.} The final value at 0 Pa and 20.0 $^\circ$C,
\begin{equation}\label{d220INRIM}
 \dd(\eSi) = 192014712.67(67)\; \rm am ,
\end{equation}
is the mean of the measurement results shown in Fig.\ \ref{comparison}. The relative uncertainty is $3.5\times 10^{-9}$. To act with caution, the uncertainty of (\ref{d220INRIM}) is the mean uncertainty of each single measurement in Fig.\ \ref{comparison}, not the value reduced by the mean.

\begin{table}[b]
\caption{\label{table:1}Relative correction and uncertainty, in parts per $10^9$, of the 10/05/05 measured $\dd$ value}
\begin{ruledtabular}
\begin{tabular}{ldd}
Contribution           &\mbox{Correction} &\mbox{Uncertainty} \\
\hline
statistics             & 0.00     & 0.36\\
wavelength             & -0.06    & 0.03\\
laser beam diffraction & 7.26     & 0.65\\
laser beam alignment   & 1.36     & 0.77\\
platform tilt          & -1.37    & 1.20\\
Abbe's error           & 0.00     & 1.50\\
trajectory             & 1.06     & 0.65\\
temperature            & -0.50    & 2.55\\
self weigh deformation & 0.81     & 0.30\\
aberration             & 0.00     & 0.50\\
\hline
total                  & 8.56     & 3.48\\
\end{tabular}
\end{ruledtabular}
\end{table}

Point defects, mainly carbon, oxygen, and vacancies, strain the crystal. The $\NA$ determination required that the defect-concentration differences between the samples used for the lattice parameter and density measurements is accounted for. Therefore, we registered the distance from the seed crystal, 299.5 mm, of the 50 mm line along which the measured $\dd$ values were averaged.

After extrapolation to an impurity free crystal, the $\eSi$ lattice parameter is larger by $1.9464(67) \times 10^{-6} a_0$ than the parameter of the natural Si crystal WASO04, similarly extrapolated to an impurity free crystal \cite{waso04}.

The dependence on isotopic composition is a combined effect of thermodynamics and quantum mechanics \cite{Pavoni:1,Biernacki}. The interatomic distance minimizes the Gibb's free energy with respect to the cell volume. In addition to the elastic energy, the free energy depends on the phonon energy as well as on the entropy associated with temperature. While the elastic energy sets an equilibrium distance independent of nuclear mass, the phonon energy does not. Anharmonic effects imply a greater equilibrium distance and they cause thermal expansion. Since heavier isotopes have smaller phonon energy, they set at a smaller distance. Entropy increases with temperature and has an opposite effect. At zero temperature, only the zero-point phonon energy survives so that $\eSi$ has the greater lattice parameter; this is a pure quantum mechanical effect. When temperature increases, the lattice parameter difference decreases, as a consequence of the increasing entropy. At room temperature, the result of quantum mechanics calculations of the lattice parameter difference, $2.03 \times 10^{-6} a_0$, between $\eSi$ and natural Si \cite{Pavoni:2} is in excellent agreement with the value we found.

We received funds from the European Community's VII Framework Programme ERA-NET Plus grant 217257.

\end{document}